\documentclass[12pt,a4paper]{article}

\usepackage{graphics}
\usepackage{latexsym}
\usepackage{amsmath}

\title{A toy model of the brain}
\author{B. Hoeneisen and F. Pasmay}
\date{\small{Universidad San Francisco de Quito} \\
	30 March 2004 \\}
\begin{document}
\maketitle

\begin{abstract}
\noindent
We have designed a toy brain and
have written computer code that simulates it.
This toy brain is flexible, modular,
has hierarchical learning and recognition, has short and long
term memory, is distributed (\textit{i.e.} has no central control), 
is asynchronous, and includes parallel and series processing.
We have simulated the neurons calculating
their internal voltages as a function of time. We include in
the simulation the ion 
pumps of the neurons, the synapses with glutamate or 
GABA neurotransmitters, and the delays of the action pulses
in axons and dendrites.
We have used known or plausible circuits of real brains.
The toy brain reads books and learns languages using
the Hebb mechanism. Finally, we have related the toy brain
with what might be occurring in a real brain. \\
\end{abstract}

\section{Introduction}
We have designed a toy brain with the model described in \cite{bh1},
and have simulated it on the computer in c++ language.
The purpose of this exercise is to simulate neurons by calculating
their internal voltage, including their
ion pumps, synapses with neurotransmitters, electrochemical
action pulses that propagate with finite speed along the
axons and dendrites, and the 
\textquotedblleft{Hebb mechanism}"\cite{neuron}
that alters the strengths of the synapses so that the brain
can learn. 
We have simulated excitatory synapses with glutamate
neurotransmitter, and inhibitory synapses with GABA 
neurotransmitter.\cite{neuron} These synapses can release
neurotransmitters fast or slow.
We have used known circuits of the brain wherever
possible,\cite{neuron} or at least circuits that the brain may plausibly
use. The toy brain can learn languages,
\textit{e.g.} English and Spanish, without mixing them
into \textquotedblleft{Spanglish}". We have designed a 
plausible structure of the brain, including parallel and
series hierarchical learning and recognition, with long
and short term memory
at the higher levels of the hierarchy.
The toy brain is flexible, modular, asynchronous, has
distributed processing (\textit{i.e.} no central control),
and its neurons have critical learning periods.
This toy brain has a little over 12 thousand neurons and
70 million synapses.

\begin{figure}
\begin{center}
\vspace*{-3.5cm}
\scalebox{0.6}
{\includegraphics[0in,1in][8in,9.5in]{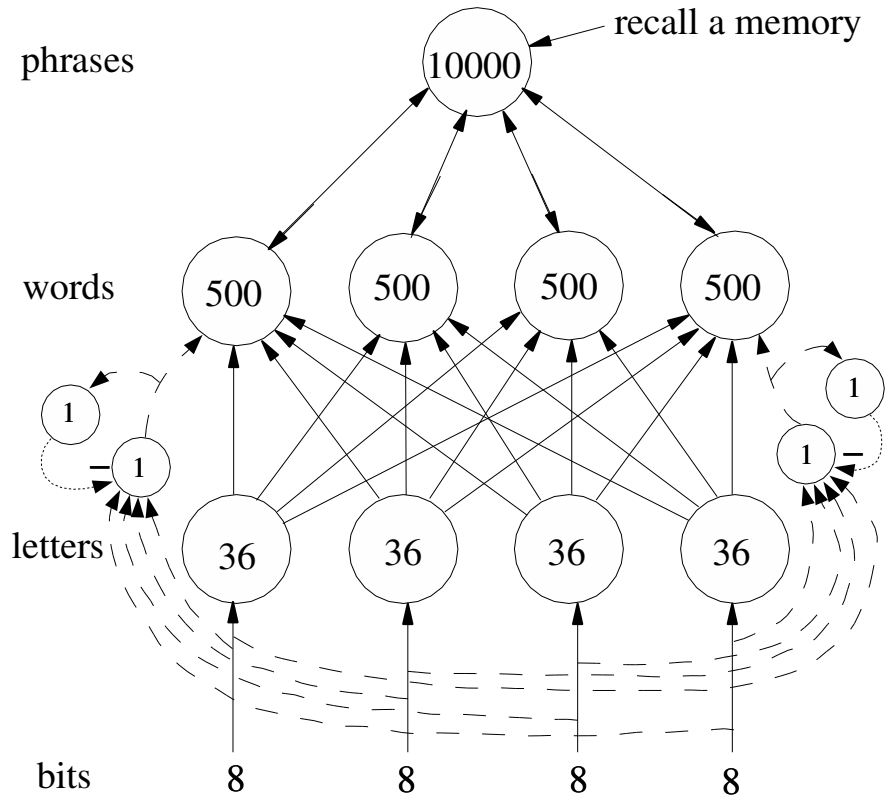}}
\vspace*{1.0cm}
\caption{Model of a toy brain with four hierarchical levels:
bits, letters, words and phrases. The number of principal neurons in 
each block are shown. The neurons in the
upper hierarchies (words and phrases)
have auxiliary neurons with slow negative feedback
to turn off oscillations after a delay. Each block of word neurons
is activated sequentially by one auxiliary neuron (shown only for
the first and last word blocks). The arrows represent 
several axon, synapse and dendrite.
A double arrow represents axons going up and axons coming down.}
\label{block}
\end{center}
\end{figure}

The block diagram of the toy brain is described in Section 2.
Then follow descriptions of the blocks, neurons, results, and the possible
relation of the toy brain with a real brain. 
Conclusions are collected at the end.

\section{Block diagram}

To be specific we will describe the particular toy brain shown
in Figure \ref{block}. 
\footnote{The numbers in Figure \ref{block} are constants of the
c++ program and can be set at will. We have tried other
brain topologies, 
but limit the discussion to the particular 
one shown in the figure.}
This toy brain can learn up to 36
letters of 8 bits, 500 words of 4 letters, and 10,000 phrases of 4
words. This brain reads a book in English, or Spanish, or whatever. 
The text is parsed into phrases of 4 words. 
Each word is truncated to the first 4 letters
(if the word is shorter than 4 letters it is padded with spaces).
Each letter is coded into 8 bits with 4 ones 
and 4 zeroes. The toy brain has 4 hierarchical levels called
\textquotedblleft{bits}", \textquotedblleft{letters}",
\textquotedblleft{words}" and \textquotedblleft{phrases}".
The \textquotedblleft{letter}" neurons learn patterns of 8 bits,
the \textquotedblleft{word}" neurons learn patterns of 4 letters, and
the \textquotedblleft{phrase}" neurons learn patterns of 4 words.

\textquotedblleft{Phrases}" and \textquotedblleft{words}" 
are accessible to short term memory,\cite{bh1}
\textit{i.e.} if a \textquotedblleft{phrase}" neuron fires it 
causes the corresponding four \textquotedblleft{word}" neurons to fire
after a delay,
which, in turn, cause the \textquotedblleft{phrase}" neuron to fire again.
Note the double arrows in Figure \ref{block}.
These synchronized firings cease after about ten seconds 
(the duration of short term memory which is settable) because 
each of these neurons has an auxiliary neuron that integrates (counts) pulses
until it reaches its threshold, and then inhibits the main neuron
with a slow inhibiting (GABA) synapsis.
Thus, according to the model of \cite{bh1}, the \textquotedblleft{phrase}" 
acquires the composite \textquotedblleft{meaning}" of its 4 \textquotedblleft{words}. 
\textquotedblleft{Letters}" and 
\textquotedblleft{bits}" are not
accessible to short term memory, \textit{i.e.} they do not have 
feedback circuits to sustain repeated firings of the neurons.
Note the single arrows in Figure \ref{block}.

The brain reads the book one word at a time. The 4 letters of the word 
are learned and recognized in parallel. To recognize a phrase of
4 words, the toy brain learns and recognizes the 4 words in series
(\textit{i.e.} one at a time). 
Each block of word neurons
is activated sequentially by one auxiliary neuron that counts 
the words (shown only for
the first and last word blocks in Figure \ref{block}). These auxiliary
neurons are reset due to the lapse of time between one phrase and the
next.

\begin{figure}
\begin{center}
\vspace*{-8.0cm}
\scalebox{0.6}
{\includegraphics[0in,1in][8in,9.5in]{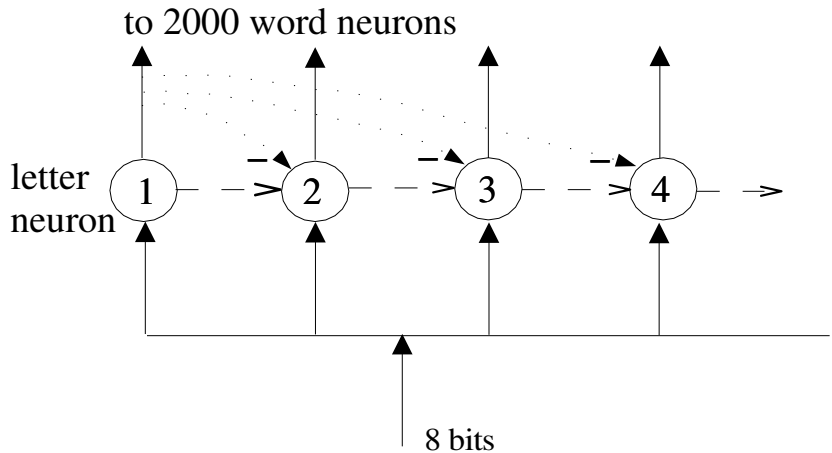}}
\vspace*{1.0cm}
\caption{Four \textquotedblleft{letter}" 
neurons in a block of 36. 
The inhibiting (GABA) synapses (shown only for neuron
\textquotedblleft{1}", see dotted lines) prevent 
repeated learning of the same letter.
The dashed lines carry the learning signal.}
\label{letter}
\end{center}
\end{figure}

\section{Blocks}
All blocks in Figure \ref{block} are similar.
Let us describe the first block with 36 \textquotedblleft{letter}" 
neurons. 
Four of these neurons are shown in Figure \ref{letter}.
Each \textquotedblleft{letter}" neuron
has (initially) 8 inputs corresponding to the 8 bits defining the
first letter of a word, and one output connected to 
2000 word neurons. 

The neurons can be in one of three
modes: inactive mode, learning mode, or recognition mode.
Initially, all neurons in the block are in inactive mode,
except for neuron \textquotedblleft{1}" which is in 
learning mode. A \textquotedblleft{letter}" with 4 ones and
4 zeroes is presented to the first block of 36 
\textquotedblleft{letter}" neurons.
Neuron \textquotedblleft{1}" fires and learns this letter by
strengthening the corresponding 4 synapses and breaking the other 4.
This is the Hebb mechanism.\cite{neuron}
Neuron \textquotedblleft{1}" passes to recognition mode, and
tells neuron \textquotedblleft{2}" to pass to learning mode.
This signal is shown with a dashed arrow in Figure \ref{letter}.
Note that neuron \textquotedblleft{2}" has reached its critical period
for learning. Before learning, a \textquotedblleft{letter}" neuron
has 8 inputs. After learning, only 4 inputs survive (with synapses
with strengthened couplings).

\begin{figure}
\begin{center}
\vspace*{-7.0cm}
\scalebox{0.6}
{\includegraphics[0in,1in][8in,9.5in]{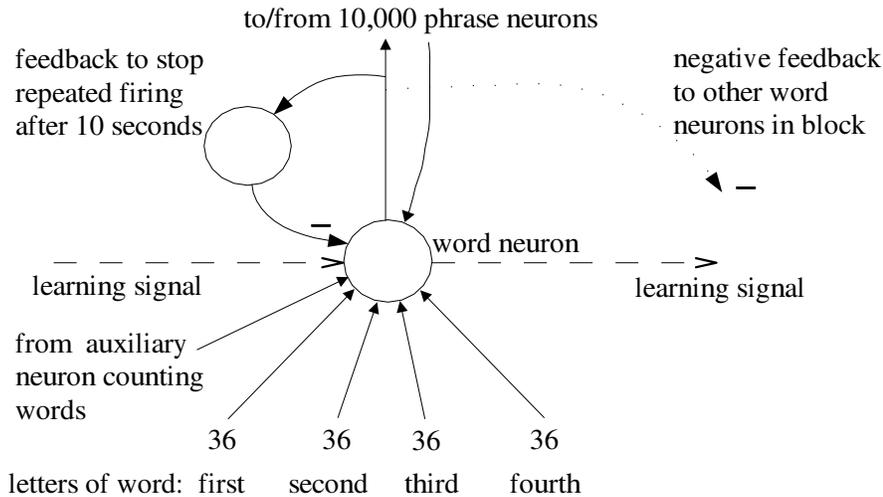}}
\vspace*{1.0cm}
\caption{One \textquotedblleft{word}" neuron with associated 
circuitry.}
\label{word}
\end{center}
\end{figure}

The output of each neuron has an inhibitory (GABA) connection to an
input of each succeeding neuron of the block.
These fast
inhibiting connections (shown only for neuron \textquotedblleft{1}"
in Figure \ref{letter}, see dotted lines) prevent
several neurons in the block from learning the same letter,
and also make the design of the brain more robust, since,
for blocks with short term memory,
only one neuron per block can fire repeatedly at any time.

Let us now consider the $n$'th \textquotedblleft{word}" 
neuron in a block
of 500 words, see Figure \ref{block}.
This neuron and associated circuitry is shown in Figure \ref{word}.
Before learning,
a \textquotedblleft{word}" neuron has $36 \cdot 4 = 144$ input
connection from \textquotedblleft{letter}" neurons. After learning,
only 4 of these connections are left (with synapses with strengthened coupling).
A \textquotedblleft{word}" neuron has the same circuits of a
\textquotedblleft{letter}" neuron, plus one added circuit.
To turn off
the repeated firings of neuron \textquotedblleft{$n$}" 
(due to the feedback loops between \textquotedblleft{phrase}" neurons
and \textquotedblleft{word}" neurons) there
is a connection from the output of neuron \textquotedblleft{$n$}"
to one of its inputs \textit{via} an auxiliary neuron. 
This auxiliary neuron has a long time constant to return to its
resting voltage, so it integrates (counts) its input pulses
until it reaches its threshold voltage. 
The output synapse of the auxiliary neuron 
is inhibitory and releases GABA neurotransmitter slowly,
so it turns off the repeated firings of neuron \textquotedblleft{$n$}"
after about 10 seconds (settable). 
Such positive and negative feedback circuits have been observed
in real brains.\cite{neuron}

The other difference between \textquotedblleft{letter}" blocks and 
\textquotedblleft{word}" blocks is that the former learn in parallel,
while the latter learn in series. To implement the series learning
of words
we need a control that activates one word block after another.
This is accomplished with one auxiliary neuron 
per block. This auxiliary neuron has a long time constant and integrates 
(counts) the words of the phrase and
fires to synchronize these words.
The first \textquotedblleft{word}" block learns the first word of
a phrase, the second block learns
the second word of the phrase, and so on. The words are synchronized
by the lapse of time between one phrase and the next.

\section{Neurons}

The simulated neurons have a resting voltage -0.06 Volt. Ion
pumps make the voltage of the neuron approach the resting
voltage with a time constant $\tau$, which is a settable
parameter. The neurons have a threshold voltage -0.04 Volt.
When the internal voltage of the neuron surpasses this
threshold, an action pulse is produced 
(due to fast sodium and slow potassium voltage activated
ion channels) with a time distribution
taken from \cite{neuron}.

Synapses are defined by the following parameters:
voltage step $\Delta V$ into the target neuron
with a settable time distribution (the injection
of charge into the target neuron can be faster or
slower), and
delay $\Delta t$ of associated dentrite and axon.
If $\Delta V > 0$, the synapse is excitatory.
If $\Delta V < 0$, the synapse is inhibitory.

Finally, if the neuron is in the learning state,
then the $\Delta V$ of all inputs that cause the neuron
to fire are increases by $\delta V$, and the
$\Delta V$ of all other inputs are set to zero
(effectively disconnecting these synapses).

Real neurons must have adjusted these parameters
(by learning from experience and/or by evolution).

\section{Results}

\begin{figure}
\begin{center}
\vspace*{-6.0cm}
\scalebox{0.7}
{\includegraphics[0in,1in][8in,9.5in]{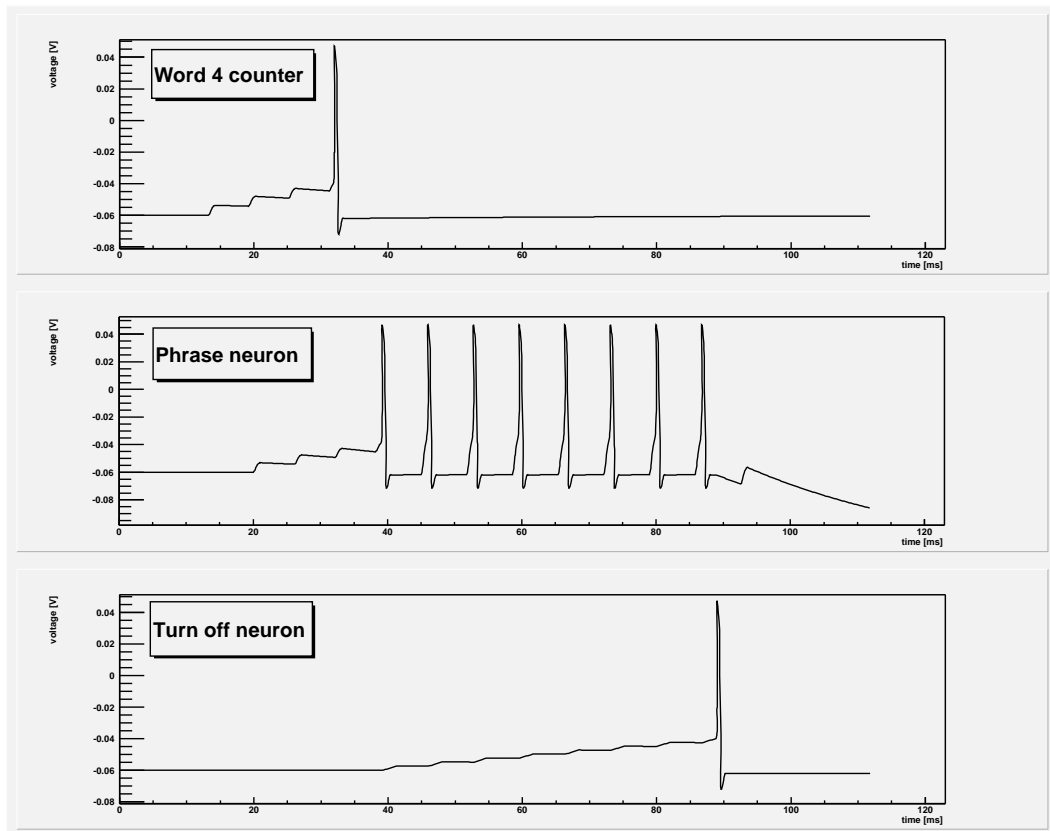}}
\vspace*{1.5cm}
\caption{Internal voltage of three neurons as a function of time.
From top to bottom: auxiliary neuron that
counts four words to enable the fourth word block;
a \textquotedblleft{phrase}"
neuron that recognizes a phrase of four words;
and the auxiliary neuron
that turns off this \textquotedblleft{phrase}" neuron after a delay.}
\label{pulses}
\end{center}
\end{figure}

The toy brain has learned (in the sense described in this note)
Spanish by reading \textquotedblleft{Vivir para contarla}" by
Gabriel Garc\'{\i}a Marquez. To show the level of detail of the simulation,
we present in Figure \ref{pulses} the internal voltage of
three neurons: the neuron that counts the fourth word in a phrase,
a phrase neuron that recognizes four words, and the
auxiliary neuron that turns this word neuron off after a delay. Note that the
neurons have a resting voltage of -0.06 Volt, and a threshold
voltage of -0.04 Volt.
Note that when the phrase neuron fires, the feedback loops cause
it to fire repeatedly, until the auxiliary shut-off neurons turn
this oscillation off. Note, in the top and middle graphs, the four
steps (corresponding to the four words) required to 
reach the threshold voltage of these neurons.

\section{Thoughts on thoughts}

\begin{figure}
\begin{center}
\vspace*{-6.5cm}
\scalebox{0.6}
{\includegraphics[0in,1in][8in,9.5in]{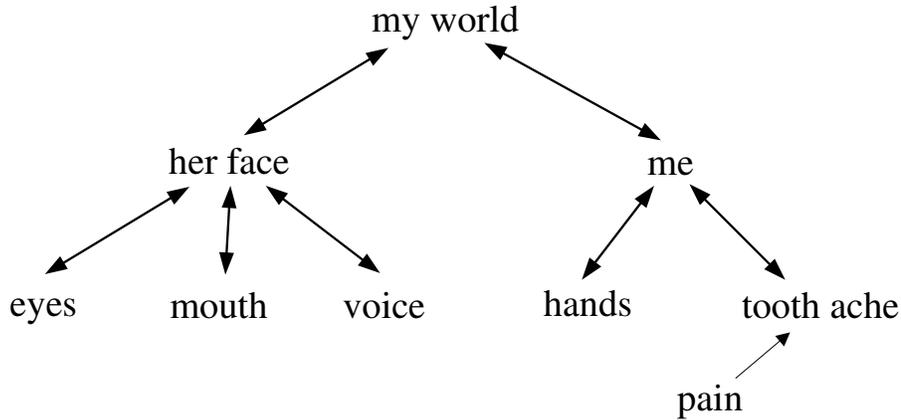}}
\vspace*{-1.0cm}
\caption{Nine neurons of a toy brain. The double
arrows represent one axon going up and one axon coming
down.}
\label{my_world}
\end{center}
\end{figure}

Let us put our toy brain in a more familiar setting.
We will add one more hierarchical level: call it 
\textquotedblleft{paragraph}".
For the sake of brevity we will consider only the nine
neurons shown in Figure \ref{my_world}: one
\textquotedblleft{paragraph}" neuron called
\textquotedblleft{my world}", two \textquotedblleft{phrase}"
neurons called \textquotedblleft{her face}" and 
\textquotedblleft{me}",
five \textquotedblleft{word}" neurons called
\textquotedblleft{eyes}", \textquotedblleft{mouth}",
\textquotedblleft{voice}", \textquotedblleft{hands}" and
\textquotedblleft{tooth ache}", and one
\textquotedblleft{letter}" neuron called \textquotedblleft{pain}".
The toy brain has many other neurons and another hierarchical
level (\textquotedblleft{bits}")
that we will not consider. The connections shown in
Figure \ref{my_world} have been learned by prior 
experience of the toy brain.

Now suppose I am discussing consciousness with a friend.
My eyes see her and my ears hear her, and my brain
recognizes her eyes, her mouth and her voice (meaning that 
the \textquotedblleft{eyes}", \textquotedblleft{mouth}" and
\textquotedblleft{voice}" neurons fire). The firing of these
neurons cause the \textquotedblleft{her face}" neuron to
fire, so I have recognized her face. Feedback loops
cause repeated firings of these neurons, binding together
those eyes, that mouth and that voice to form her face.

In a real brain, the neurons in the upper hierarchies 
are associated with awareness. 
We assume that \textquotedblleft{bits}" neurons have no awareness,
\textquotedblleft{letter}" neurons have awareness and are not accessible
from short term memory (see the single ended arrow in Figure \ref{my_world}),
and \textquotedblleft{word}", \textquotedblleft{phrase}" and
\textquotedblleft{paragraph}" neurons have awareness and are
accessible from short term memory. 
So now, in a real brain,
I am aware of that face with those eyes, that mouth and that voice.

My eyes also see my hands. My brain recognizes these hands, 
meaning that the \textquotedblleft{hands}" neuron fires. Also,
a neuron in my tooth causes the \textquotedblleft{pain}" neuron
to fire, which in turn causes the
\textquotedblleft{tooth ache}"
neuron to fire. The firing of the
\textquotedblleft{hands}" neuron and the 
\textquotedblleft{tooth ache}" neuron cause the
\textquotedblleft{me}" neuron to fire, so I have recognized
my self. Feedback loops cause repeated firings of these neurons,
tying me together, so now I am aware of my self with these
hands and this tooth ache with that pain.

The firing of the \textquotedblleft{her face}" and 
\textquotedblleft{me}" neurons in turn cause the firing of the
\textquotedblleft{my world}" neuron. Feedback loops now cause
eight neurons to fire synchronously,
tying my world together, so now I
become aware of my world with her face (with those eyes, that mouth
and that voice) and me (with these hands and this tooth ache
with that pain).
The \textquotedblleft{pain}" neuron is not part of the feedback loops,
so the pain goes away as soon as the dentist lifts that awful
hook.

Of course this model and description are over simplified, but they give
a general idea of what might be going on in a real brain. 
Laking an understanding of the three difficult phenomena:
awareness, attention and free will (if indeed we have free will!),
we present the following speculations as a working hypothesis
to help guide experiments. 

\textbf{Hypothesis:} From a certain hierarchical level on
up wards:
\begin{enumerate}
\item
Awareness is the dynamic set of firing neurons.
\item
Somehow, I can excite a neuron (by free will?).
\item
Somehow, I can inhibit all neurons belonging to a hierarchical
level (by free will?).
\end{enumerate}
If (by Hypothesis 2)
I excite, for example, the neuron \textquotedblleft{her face}", then
feedback loops make \textquotedblleft{eyes}", \textquotedblleft{mouth}",
\textquotedblleft{voice}" and \textquotedblleft{her face}" to
fire repeatedly in synchronism. In this way I have brought
the image of that face (with those eyes, that mouth and that voice)
from long term memory and have re-lived it in
short term memory.
Note that if I excite the \textquotedblleft{me}" neuron I can
remember myself with these hands and this tooth ache, but I can 
(fortunately) not feel the pain of that
tooth ache from memory (because the \textquotedblleft{pain}"
neuron is not within the feedback loops and can not be excited
from memory). 

To focus attention on, for example, her eyes, I must inhibit all 
\textquotedblleft{word}" neurons except \textquotedblleft{eyes}".
A way to do that is to inhibit all \textquotedblleft{word}" neurons
(Hypothesis 3) and excite the \textquotedblleft{eyes}" 
neuron (Hypothesis 2).

Ren\'{e} Descartes divided the world into \textit{res extensa} (matter) and
\textit{res cogitans} (mind). The problem with this division 
is that it leaves out the interaction of matter on mind, and mind on
matter. May we suggest identifying \textquotedblleft{mind}" with the
\textquotedblleft{dynamic set of firing neurons}" (in the upper hierarchical
levels of processing). Then the interaction of matter on mind
(\textit{via} synapsis from sense cells to neurons), and mind on matter
(\textit{via} synapsis from neurons to muscles) becomes clear. 

Let us repeat Hypothesis 1: Awareness is the dynamic set of firing neurons
(in the upper hierarchical levels of processing). This set of
firing neurons is determined by inputs from the senses and/or memory
(Hypothesis 2), and/or attention (Hypothesis 2 and 3), and a lifelong
learning process that programmed the strengths of the
couplings at the synapses (and other parameters of neurons
and synapses), and is partially
\footnote{In our example, the
\textquotedblleft{pain}" neuron is outside of the feedback loops.}
sustained for a few seconds by feedback loops.
According to Hypothesis 1, we have identified what I am aware of
with the set of firing neurons.
The pain of the tooth is a set of firing neurons.
Who/what feels that pain: a grander set of firing neurons.
The sight of her face is a pattern of firing neurons.
Who/what sees her face: a grander set of firing neurons.
The awareness of my world is a set of firing neurons. 
What/who has that awareness: that same pattern of firing neurons.
If we define \textquotedblleft{I}" as what I am aware of, then
\textquotedblleft{I}" am a dynamic set of firing neurons,
sustained by the organization of my brain, the organization of
my body, and by the world.

Our toy brain does not understand what it reads, for the same reason I
do not understand a book in Polish
(I would have understood the book if my mother would have been
Polish).
For the toy brain to understand,
it is necessary (but hardly sufficient)
to equip it with sensors and hands so it can explore
the world, and give it a \textquotedblleft{mother}" to teach 
it the names of the things in the world. Would the toy brain then
wake up to awareness?

\section{Conclusions}

We have designed and developed a toy brain that runs on 
a personal computer. 
This toy brain is flexible, modular,
has hierarchical learning and recognition, has short and long
term memory, is distributed (\textit{i.e.} has no central control), 
is asynchronous, and includes parallel and series processing.
We have simulated the neurons calculating
their internal voltages as a function of time. We include in
the simulation the ion 
pumps of the neurons, the synapses with glutamate or 
GABA neurotransmitters, and the delays of the action pulses
in axons and dendrites.
We have used known or plausible circuits of real brains.
The toy brain reads books and learns languages 
(in the limited sense described in the article)
using the Hebb mechanism. We have identified parameters that real
neurons must be able to adjust (by learning from experience
and/or evolution). 
We have related the toy brain
with what might be occurring in a real brain, and have
obtained some insight on our minds.
We propose that neurons in a real brain
can be in inactive mode, learning mode or recognition mode.
Finally we have presented working hypothesis related to
the difficult questions: awareness, attention and free will.

\end{document}